\documentclass[12pt,dvips]{article}

\usepackage{rotating}
\usepackage{epsfig}
\usepackage{amssymb}
\usepackage{graphicx}
\usepackage{pdproc}

  \makeatletter
  \def\@cite#1{[#1]}
  \makeatother
\begin{document}

\renewcommand{\thefootnote}{\alph{footnote}}

\title{Determining SUSY and Higgs Parameters in the MSSM and its Extensions}

\author{SEONG YOUL CHOI}

\address{Department of Physics, Chonbuk National University, Chonju 561--756,
         Korea\\ {\rm E-mail: sychoi@chonbuk.ac.kr}}

\abstract{If supersymmetry (SUSY) is realized at the electroweak scale, its
          underlying structure and breaking mechanism may be explored with
          great precision by a future linear $e^+ e^-$ collider (LC) with a
          clean environment, tunable collision energy, high luminosity polarized
          beams, and additional $e^-e^-$, $e\gamma$ and $\gamma\gamma$ modes.
          We review a few recent developments for
          determining fundamental SUSY and Higgs parameters, measuring CP
          violating $H/A$ mixing in the decoupling regime and probing the
          next--to--minimal supersymmetric standard model at the LC.}

\normalsize\baselineskip=15pt
\vskip 0.4cm

\section{Introduction}
\label{sec:introduction}

Weak--scale SUSY has its natural solution to the gauge hierarchy problem,
providing a stable bridge between the electroweak scale and the grand
unification or Planck scale, with which the roots of standard particle physics
are expected to go as deep as the Planck length of $10^{-33}$\,cm. It is then
crucial to probe SUSY and its breaking with great precision at a future
$e^+e^-$ linear collider (LC) \cite{LC} as well as the large hadron
collider (LHC) \cite{LHC} for a reliable grand extrapolation
to the Planck scale \cite{RGE}.

In this talk we review a few recent developments for determining fundamental
SUSY and Higgs parameters, measuring CP violating $H/A$ mixing in the decoupling
regime and probing the next--to--minimal supersymmetric standard model (NMSSM)
at the LC.

\section{A new $\tan\beta$ determination method: $\tau\tau$ fusion to SUSY Higgs
         bosons}

For large pseudoscalar Higgs  mass the heavy $H/A$ Higgs couplings to
down-type fermions are directly proportional to $\tan\beta$ if the parameter
is large so that they are highly sensitive to its value \cite{R2}.
Also the down-type couplings of the light $h$ Higgs boson in the MSSM are close
to $\tan\beta$ if $M_A$ is moderately small. Based on these
observations, we show that $\tau\tau$ fusion to Higgs bosons at a photon
collider \cite{2a} can provide a valuable method for measuring $\tan\beta$, after
searching for Higgs bosons in $\gamma\gamma$ fusion \cite{2b}.

For large $\tan\beta$, all the Higgs bosons $\Phi$ ($=H, A, h$) decay almost
exclusively [80 to 90\%] to a pair of $b$ quarks so that the final state consists
of a pair of $\tau$'s and a pair of resonant $b$ quark jets. Two main background
processes - the $\tau^+ \tau^-$ annihilation into a pair of $b$-quarks via
$s$-channel $\gamma/Z$ exchanges and the diffractive $\gamma \gamma \to
(\tau^+ \tau^-) (b\bar{b})$ events with the pairs scattering off each other by
Rutherford photon exchange - can be suppressed strongly by choosing proper
cuts \cite{2a}.

\begin{figure}[!htb]
\begin{center}
\includegraphics*[width=5cm, height=4cm]{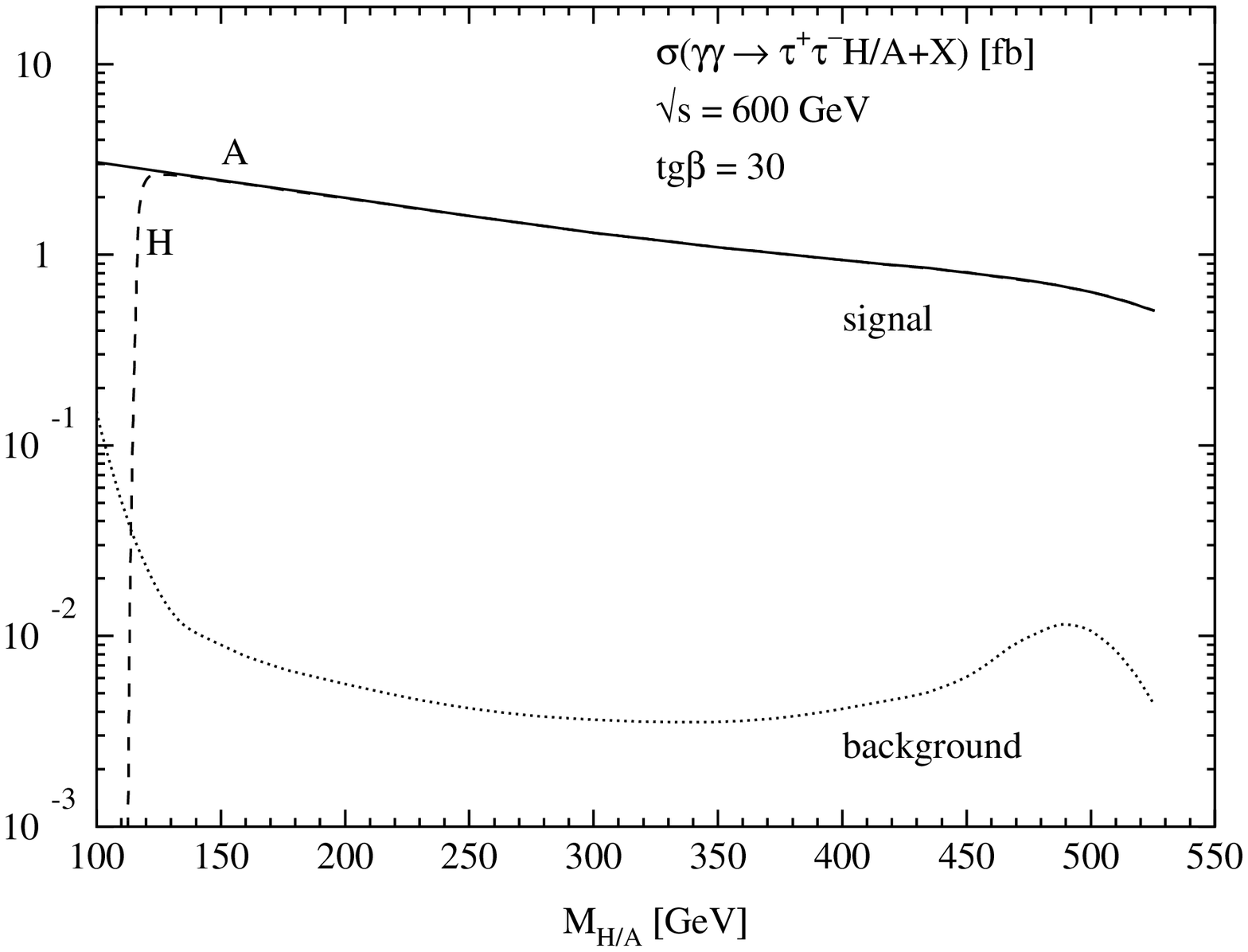}
\hskip 1.cm
\includegraphics*[width=5cm, height=4cm]{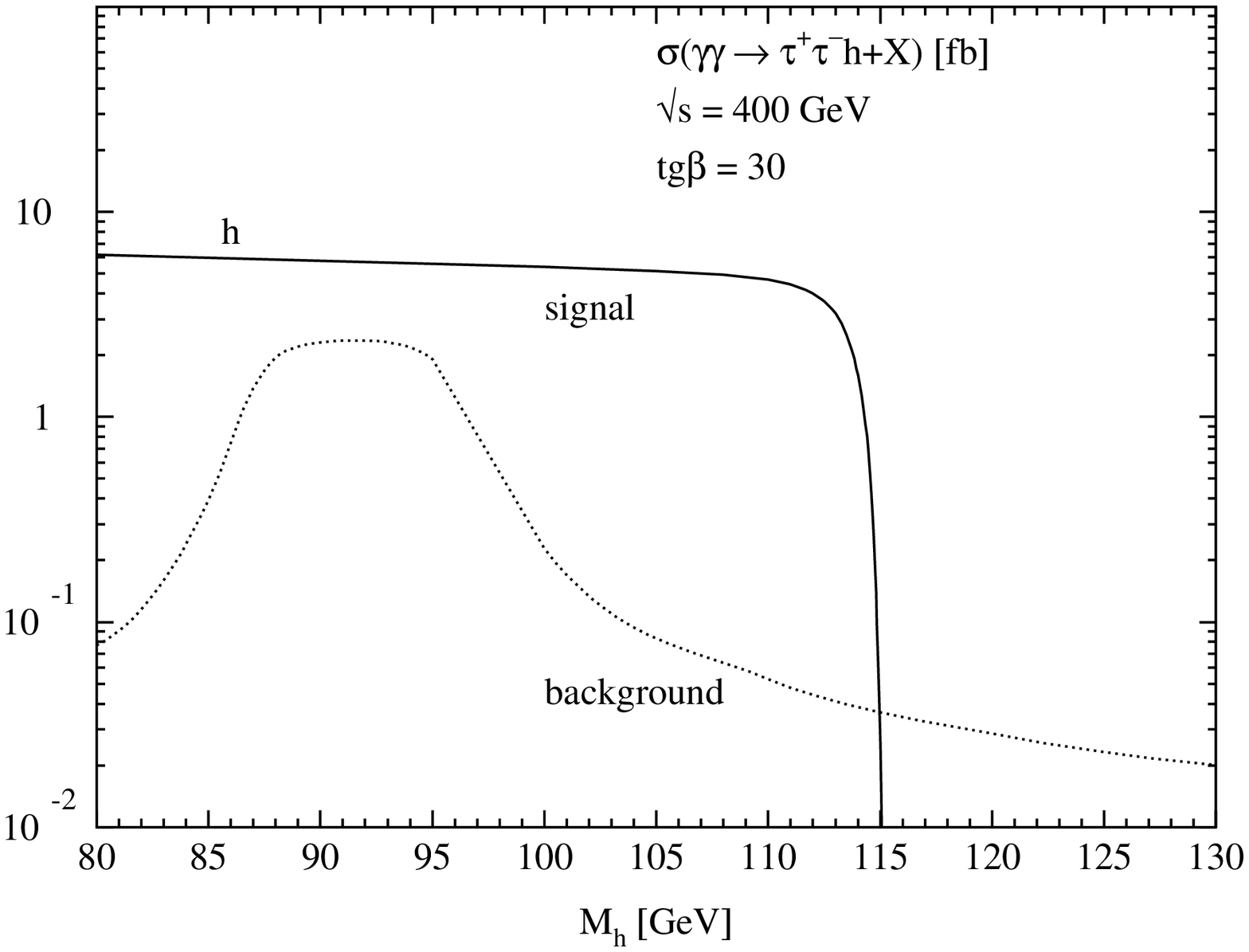}
\caption{\it The cross sections for the production of the $H/A$ (left)
  and $h$ (right) Higgs bosons in the $\tau\tau$ fusion process
  at a $\gamma\gamma$ collider for
  $\tan\beta=30$.  Also shown is the background cross section with experimental
  cuts. $\sqrt{s}$ denotes the $\gamma\gamma$ collider c.m. energy.}
\label{fig:xsec}
\end{center}
\end{figure}

The left panel of Fig.~\ref{fig:xsec} shows the exact cross sections for
the signals of $H$ and $A$ Higgs-boson production in the $\tau\tau$ fusion
process with $E_{\gamma\gamma}=600$ GeV, together with all the background
processes with appropriate experimental cuts. As shown in the right panel of
Fig.~\ref{fig:xsec} $\tau\tau$ fusion to the light Higgs boson $h$ with
$E_{\gamma\gamma}=400$ GeV can also be exploited to measure large $\tan\beta$ for
moderately small $M_A$.
For $h$ production, the mass parameters are set to $M_A\sim 100$ GeV and
$M_h=100$ GeV. The channels $h/A$ and $H/A$ are combined in the overlapping mass
ranges in which the respective two states cannot be discriminated. Since in the
region of interest the $\tau\tau$ fusion cross sections are proportional to
$\tan^2\beta$ and the background is small, the absolute errors $\Delta\tan\beta$
are nearly independent of $\tan\beta$, varying between $\sim 0.9$ and $1.3$
for Higgs masses away from the kinematical limits for the integrated luminosity
of 200/100 fb$^{-1}$ for the high/low energy option.

\section{Probing Majorana nature and CP violation in the neutralino system}

Once several neutralino candidates are observed it will be crucial to
establish the Majorana nature and CP properties of neutralinos
as well as to reconstruct the fundamental SUSY parameters at the LC \cite{REC}.
In this report, we present two powerful methods for probing the Majorana nature
and CP violation in the neutralino system.

When the electron/fermion masses are neglected both the
production processes, $e^+e^- \to \tilde{\chi}^0_i\tilde{\chi}^0_j$,
near threshold and the three--body decays, $\tilde{\chi}^0_i \to \tilde{\chi}^0_j
f\bar{f}$, near the fermion invariant mass end point are effectively
regarded as processes of a static (axial--)vector current
exchange between two neutralinos. In the CP invariant case, the
neutralino $\{ij\}$ pair production and the decay $\tilde{\chi}^0_i \to
\tilde{\chi}^0_j\, V$ through a vector current satisfy the CP relations
\begin{eqnarray}
1\, =\, \pm\eta^i \eta^j\, \left(-1\right)^L
\label{eq:selection}
\end{eqnarray}
for static neutralinos, with $\eta^i=\pm i$ the intrinsic $\tilde{\chi}^0_i$ CP
parity and $L$ the orbital angular momentum of the produced pair $\{ij\}$ and of
the final state of $\tilde{\chi}^0_j$ and $V$, respectively. Therefore, in the CP
invariant case, if the production of a pair of neutralinos with the same (opposite)
CP parity  is excited slowly in $P$ waves (steeply in $S$ waves), then the
neutralino to neutralino transition is excited sharply in $S$ waves (slowly in $P$
waves).

In the CP noninvariant case the orbital angular momentum is no longer restricted
by the selection rules (\ref{eq:selection}). Consequently, CP violation in the
neutralino system can clearly be signalled by (a) the sharp $S$--wave excitations
of the production of three non--diagonal $\{ij\}$, $\{ik\}$ and $\{jk\}$ pairs
near threshold \cite{REC,Threshold1} or by (b) the simultaneous $S$--wave
excitations of the production of any non--diagonal $\{ij\}$ pair in $e^+e^-$
annihilation near threshold and of the fermion invariant mass distribution of the
neutralino three--body decays $\tilde{\chi}^0_i \to \tilde{\chi}^0_j f\bar{f}$
near the kinematical end point \cite{Threshold2}. Note that even the combined
analysis of the production of the lighter neutralino $\{12\}$ pair and the
associated decay $\tilde{\chi}^0_2 \to \tilde{\chi}^0_1 f\bar{f}$ enables us to
probe CP violation in the neutralino system.

Once two--body neutralino decays are open,
the combined production--decay analysis cannot be exploited for probing CP
violation in the neutralino system. Nevertheless, if the two--body
decays $\tilde{\chi}^0_i \to \tilde{\chi}^0_j Z$ is not too
strongly suppressed, the $Z$ polarization reconstructed via
leptonic $Z$--boson decays with great precision allows us to probe
the Majorana nature and CP violation in the neutralino system
\cite{Twobody}.

\section{Resonant CP violating $H/A$  mixing in the decoupling regime}

With non--vanishing CP phases in the soft SUSY--breaking terms in the MSSM,
radiative corrections induce two CP--even Higgs bosons, $h$ and $H$, and one
CP--odd Higgs boson, $A$, to mix forming a triplet $(H_1,H_2,H_3)$
without definite CP parities \cite{Higgs_CP}. The $H/A$ mixing can be
large in the limit with heavy and nearly--degenerate $H$ and $A$.
The lightest Higgs $H_1$ then becomes the  SM--like Higgs,
and does not mix with the $H/A$ system.

For small mass differences, the mixing is strongly affected by the
widths of the states and the complex, symmetric Weisskopf--Wigner
mass matrix ${\cal M}^2_c=M^2-iM\Gamma$ must be considered in total, not
only the real part. Recently a coupled-channel method has been
employed \cite{ELP} for the Higgs formation and decay processes at the LHC.
We have presented an alternative approach in Ref.\cite{C1} where
the full mass matrix ${\cal M}^2_c$ is diagonalized by a complex rotation
matrix. For the $H/A$ system, the 2$\times$2 rotation matrix
is expressed in terms of a {\it complex} mixing angle $\theta$, satisfying
\begin{eqnarray}
X = (1/2)\tan2\theta={\cal M}^2_{HA}/({\cal M}^2_{HH}-{\cal M}^2_{AA})
\end{eqnarray}
where ${\cal M}^2_{HA}$ and ${\cal M}^2_{HH,AA}$ are the off--diagonal and
diagonal entries of the matrix ${\cal M}^2_c$.

\begin{figure}[!htb]
\begin{center}
\includegraphics*[width=4.5cm,height=4cm]{reimx_mssm.eps}
\hskip 0.5cm
\includegraphics*[width=9cm,height=4cm]{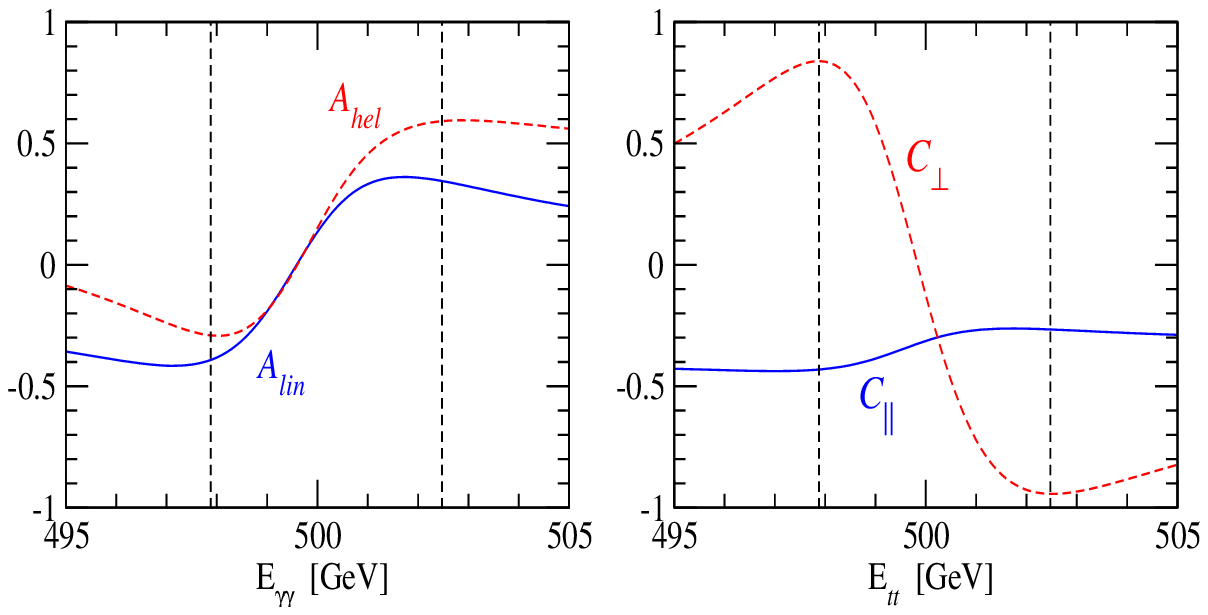}
\caption{Left: The $\phi_A$ dependence of the parameter $X$
    (left). Middle: The $E_{\gamma\gamma}$ dependence of ${\cal A}_{lin,hel}$
    in the process $\gamma\gamma \rightarrow H_i$. Right:  The
    $E_{t\bar{t}}$  dependence  of ${\cal C}_{\parallel,\perp}$ in the
    production--decay chain $\gamma\gamma \rightarrow H_i\rightarrow
    t\bar{t}$. In the middle and right figures $\phi_A=3\pi/4$. The vertical
    lines represent the two mass eigenstates.}
\label{fig:arg}
\end{center}
\end{figure}

The complex $H/A$ mixing  is shown in the left panel of
Fig.~\ref{fig:arg} for $M_S$=0.5 TeV, $|A_t|$= 1 TeV, and
$\mu$=1 TeV, $\tan\beta$=5, while varying the phase $\phi_A$ of
the trilinear parameter $A_t$.

A photon linear collider would be an ideal tool to study resonant
CP violation in the Higgs sector. Two promising signatures have
been considered in Ref.\cite{C1}. For linearly polarized photons, the
CP--even (CP--odd) component of the $H_i$ wave-functions is projected out
if the polarization vectors are parallel (perpendicular), respectively.
This can be observed in the CP--even asymmetry
${\cal A}_{lin}$, since $|{\cal A}_{lin}|$$<$1 requires both scalar and
pseudoscalar $\gamma\gamma H_i$ non-zero couplings.
Moreover, CP violation due to $H/A$ mixing can directly be probed via
the CP--odd asymmetry ${\cal A}_{hel}$ constructed with circular photon
polarization. In addition, correlations between the transverse $t$
and $\bar{t}$ polarization vectors $s_{\bot},\bar{s}_{\bot}$ in the decay
process $H_i\rightarrow t\bar{t}$, lead to a non--trivial CP-even correlation
${\cal C}_\parallel =\left\langle s_\perp \cdot \bar{s}_\perp \right\rangle$
and a CP-odd azimuthal correlation ${\cal C}_\perp = \left\langle \hat{p}_t\cdot
(s_\perp\times\bar{s}_\perp)\right\rangle$.

The middle panel of Fig.~\ref{fig:arg} shows the asymmetries ${\cal
A}_{lin}$ (solid line) and ${\cal A}_{hel}$ (dashed line)
in the $\gamma\gamma $ collider as the $\gamma\gamma$ energy is
scanned from below $M_{H_3}$ to above $M_{H_2}$. The right panel
shows the $E_{t\bar{t}}$ dependence of the correlators ${\cal
C}_\parallel$ (solid line) and ${\cal C}_\perp$ (dashed
line) for $\phi_A=3\pi/4$, a phase value close
to resonant CP mixing. Detailed experimental simulations would be needed to
estimate the accuracy with which they can be measured. However, the large
magnitude and the rapid, significant variation of the asymmetries through the
resonance region would be a very interesting effect to observe in any case.

\section{Neutralino sector in the next--to--minimal supersymmetric
         standard model}

The NMSSM superpotential \cite{Miller:2003ay,Choi:2004zx} with
an iso--singlet Higgs superfield $\hat{S}$ in addition to the two Higgs doublets
superfields $\hat{H}_{u,d}$ is given by
\begin{eqnarray}
W=W_Y +\lambda \hat{S}(\hat{H}_u \hat{H}_d)+\frac{1}{3}\kappa\hat{S}^3
\label{eq:superpotential}
\end{eqnarray}
where $W_Y$ denotes the usual MSSM Yukawa components. The two dimensionless
parameters $\lambda$ and $\kappa$ are less than 0.7 and $\kappa\lesssim\lambda$
is favored at the electroweak scale if they remain weakly interacting up to
the GUT scale \cite{Miller:2003ay}.

\begin{figure}[!htb]
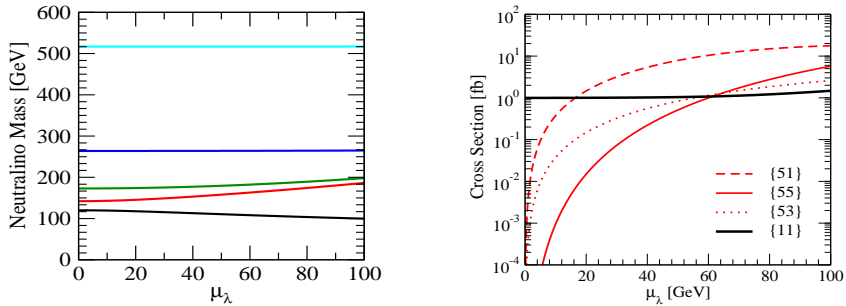

\begin{center}
\includegraphics*[width=5cm,height=4cm]{nmass_ml.eps}
\hskip 1.cm
\includegraphics*[width=5cm,height=4cm]{prod.eps}
\caption{The neutralino mass spectrum (left) and the cross-sections for
         $e^+e^- \to \tilde \chi_i^0 \tilde\chi_j^0$ at $\sqrt{s}=500$ GeV
         (right) as a function of $\mu_{\lambda}$, for $\mu_{\kappa}=120$ GeV,
         $\mu=170$ GeV, $\tan \beta=3$, $M_{1,2}=250/500$ GeV.}
     \label{fig:neut}
\end{center}
\end{figure}

The singlet superfield adds an extra higgsino to the MSSM neutralino
spectrum, called a {\it singlino}, resulting in five neutralinos. We denote
the singlino dominated neutralino $\tilde \chi_5^0$, with $\tilde \chi_{1-4}^0$
denoting the other four neutralinos in order of ascending mass. The neutralino
spectrum for an example scenario is shown in Fig.~\ref{fig:neut} (left) as a
function of $\mu_{\lambda} \equiv \lambda v/\sqrt{2}$.

In this scenario, the singlino dominated neutralino (black) is the lightest
neutralino (and the LSP) with a mass of approximately $\mu_{\kappa} \equiv 2
\kappa \langle S \rangle$ so that it will be copiously produced at the LHC in
squark and gluino cascade decays. A very decoupled state with low $\lambda$ can
give rise to macroscopic flight distances of order a $\mu$m and order a nm
for the decays $\tilde \chi_1^0 \to \tilde \chi_5^0 l^+l^-$ and
$\tilde l_R \to \tilde \chi_5^0 l$ with $\mu_{\lambda}=1$~GeV, respectively.
Also shown in Fig.~\ref{fig:neut} (right) are the cross sections for $e^+e^-
\to \tilde \chi_i^0 \tilde \chi_j^0$, for production of singlino-like
($\tilde \chi_5^0$), gaugino-like ($\tilde \chi_1^0$) and higgsino-like
($\tilde \chi_3^0$) neutralinos. With the integrated luminosity of
$1 \; {\rm ab}^{-1}$, large event rates of order $10^3$ are expected
unless $\mu_{\lambda}$ is too small.

For $\kappa \gtrsim \lambda/2$, the singlino $\tilde{\chi}^0_5$ is no longer the
LSP and it can decay to $\tilde \chi_1^0$. Such a neutralino sector would be very
difficult to distinguish from that of the MSSM.

\section{Acknowledgements}

Thanks go to J. Kalinowski, Y.G. Kim, Y. Liao, J.S. Lee, D.J. Miller,
G. Moortgat--Pick, M. M\"{u}hlleitner, M. Spira and P.M. Zerwas for fruitful
collaborations.
The work was supported by KOSEF through CHEP at Kyungpook
National University.

\bibliographystyle{plain}

\end{document}